\documentclass[doc]{apa6}
\usepackage[american]{babel}

\usepackage{csquotes}
\renewcommand*{\thefootnote}{\fnsymbol{footnote}}

\title{On the emergence of syntactic structures: \\quantifying and
  modelling duality of patterning\protect\footnote{to appear in the \emph{Proceedings of the Workshop on Origins of Communication Systems: Modeling and Ethologically-based Theory, Konrad Lorenz Institute for Evolution and Cognition Research, Altenberg, Austria (2013). Special Issue of topiCS on New frontiers in language evolution and development.}}} \author{Vittorio Loreto$^{1,2}$,
  Pietro Gravino$^{3,1}$, \\Vito D.P. Servedio$^{1}$ \& Francesca
  Tria$^2$}

\shorttitle{On the emergence of syntactic structures...}

\affiliation{
  $^1$Sapienza University of Rome, Physics Dept., Piazzale  Aldo Moro 2, 00185, Roma, Italy\\
  $^2$Institute for Scientific Interchange ISI, Via Alassio 11/c, 10126 Torino, Italy\\
  $^3$University of Bologna, Physics Dept., Via Irnerio 46, Bologna,
  Italy}

\leftheader{Loreto, Gravino, Servedio \& Tria}

\abstract{The complex organization of syntax in hierarchical
  structures is one of the core design features of human language.
  Duality of patterning refers for instance to the organization of the
  meaningful elements in a language at two distinct levels: a
  combinatorial level where meaningless forms are combined into
  meaningful forms and a compositional level where meaningful forms
  are composed into larger lexical units. The question remains wide
  open regarding how such a structure could have emerged. Furthermore
  a clear mathematical framework to quantify this phenomenon is still
  lacking. The aim of this paper is that of addressing these two
  aspects in a self-consistent way. First, we introduce suitable
  measures to quantify the level of combinatoriality and
  compositionality in a language, and present a framework to estimate
  these observables in human natural languages. Second, we show that
  the theoretical predictions of a multi-agents modeling scheme,
  namely the Blending Game, are in surprisingly good agreement with
  empirical data. In the Blending Game a population of individuals
  plays language games aiming at success in communication. It is
  remarkable that the two sides of duality of patterning emerge
  simultaneously as a consequence of a pure cultural dynamics in a
  simulated environment that contains meaningful relations, provided a
  simple constraint on message transmission fidelity is also
  considered}


 \usepackage{graphicx}
 \usepackage{url}

\begin{document}

\maketitle

\renewcommand*{\thefootnote}{\arabic{footnote}}
\setcounter{footnote}{0}

\section{Introduction}

In a seminal paper, Charles Hockett~\cite{hockett_1960_SA} identified
duality of patterning as one of the core design features of human
language. A language exhibits duality of patterning when it is
organized at two distinct levels. At a first level, meaningless forms
(typically referred to as phonemes) are combined into meaningful units
(henceforth this property will be referred to as {\em
  combinatoriality}). For example, the English forms /k/, /a/, and /t/
are combined in different ways to obtain the three words /kat/, /akt/,
and /tak/ (respectively written 'cat', 'act' and 'tack'). Because the
individual forms in them are meaningless, these words have no relation
in meaning in spite of being made of the same forms. This is a very
important property, thanks to which all of the many words of the
English lexicon can be obtained by relatively simple combinations of
about forty phonemes. If phonemes had individual meaning, this degree
of compactness would not be possible. At a second level, meaningful
units (typically referred to as morphemes) are composed into larger
units, the meaning of which is related to the individual meaning of
the composing units (henceforth this property will be referred to as
{\em compositionality}). For example, the meaning of the word
'boyfriend' is related to the meaning of the words 'boy' and 'friend'
which composed it. The compositional level includes syntax as well.
For example, the meaning of the sentence 'cats eat fishes' is related
to the meaning of the words 'cats', 'eat', and 'fishes'. In this
paper, for the sake of simplicity, we focus exclusively on the lexicon
level. This has to be considered as a first step towards the
comprehension of the emergence of complex structures in languages.

\section{Quantifying duality of patterning}

In this section we quantify the notion of duality of patterning as
observed in real languages in order to provide suitable measures for
the combinatoriality and compositionality.

\paragraph{Datasets} To this end we turned our attention to a set of
dictionaries of real languages. In order to embrace the largest
possible lists of words, i.e., including proper names and
morphological derivations, we focused our attention on the list of
words used by the Ubuntu spell checker. These lists are freely
available on the standard repositories of Ubuntu, available at the
standard Ubuntu repositories. We choose in particular British English,
French, German, Italian and Spanish\footnote{Downlodable at
  http://www.debian.org/distrib/packages}.

In addition to the above mentioned lexica we considered the list of
words produced in \emph{Human Brain Cloud} (HBC)\footnote{Dataset
  courtesy of Kyle Gabler, \url{http://kylegabler.com/}}, a massively
multiplayer English-based word association game. HBC represents the
largest available word-association database, consisting of of $78,954$
words and $1,474,006$ associations, considerably larger than any
previous word-association experiments (see~\cite{hbc} for a detailed
discussion). An interesting point of view is to look at the ensemble
of words and associations as a complex network, where nodes are words
and links are associations, and to analyze it as a word-association
graph.

\paragraph{The phonetic transcription}

Our analysis of combinatoriality and compositionality has been
performed by subdividing each word of the lexicon in its phonetic
components according to the International Phonetic
Association\footnote{http://www.langsci.ucl.ac.uk/ipa} (IPA) coding.
In order to perform easily an automated we adopted a particular
encoding of the IPA alphabet, namely the Speech Assessment Methods
Phonetic
Alphabet\footnote{http://www.phon.ucl.ac.uk/home/sampa/index.html}
(SAMPA), a computer-readable phonetic script using ASCII characters.
More in details we used eSpeak, a compact open source software speech
synthesizer that translates any texts in its SAMPA encoding.

\paragraph{Phoneme statistics}

\begin{figure}
\begin{centering}
\includegraphics[width=0.32\textwidth]{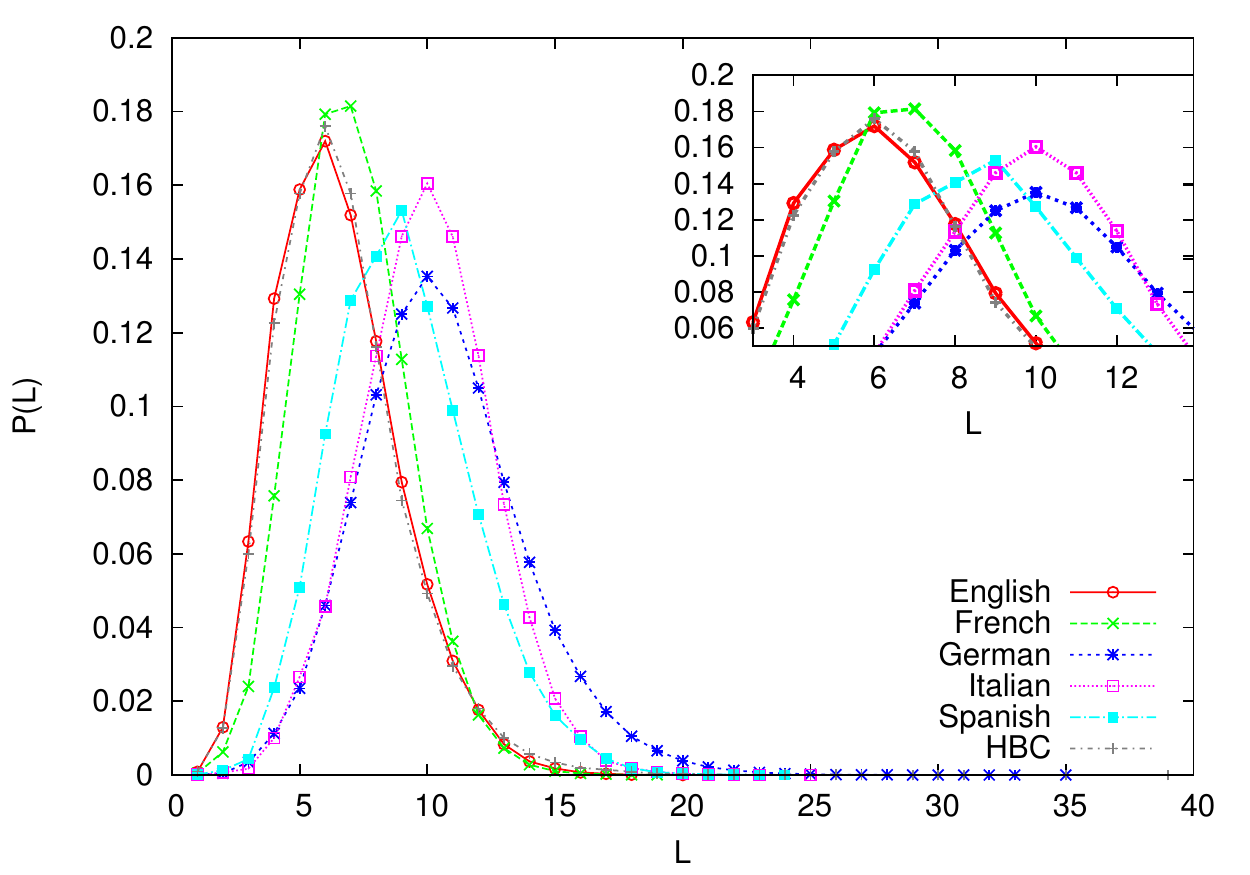}
\includegraphics[width=0.32\textwidth]{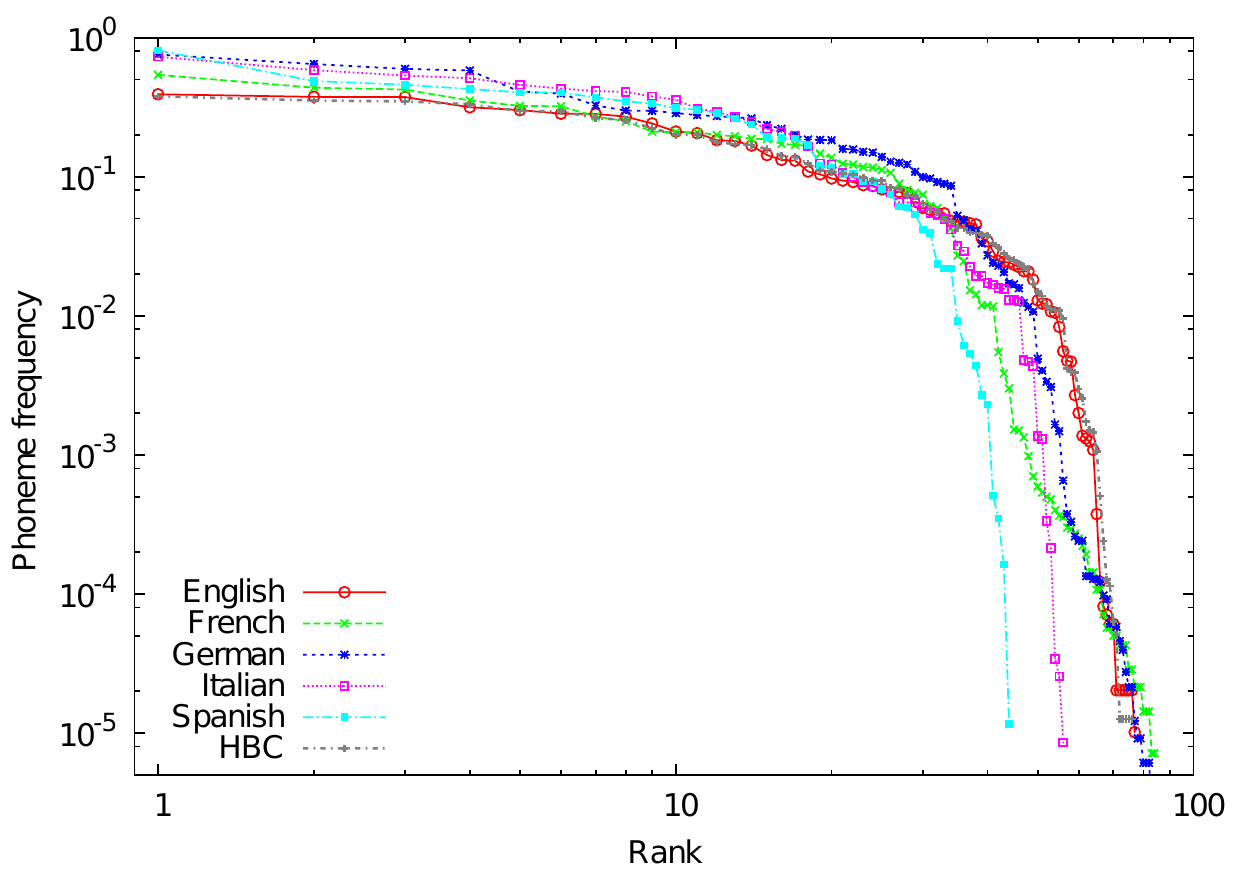}
\includegraphics[width=0.32\textwidth]{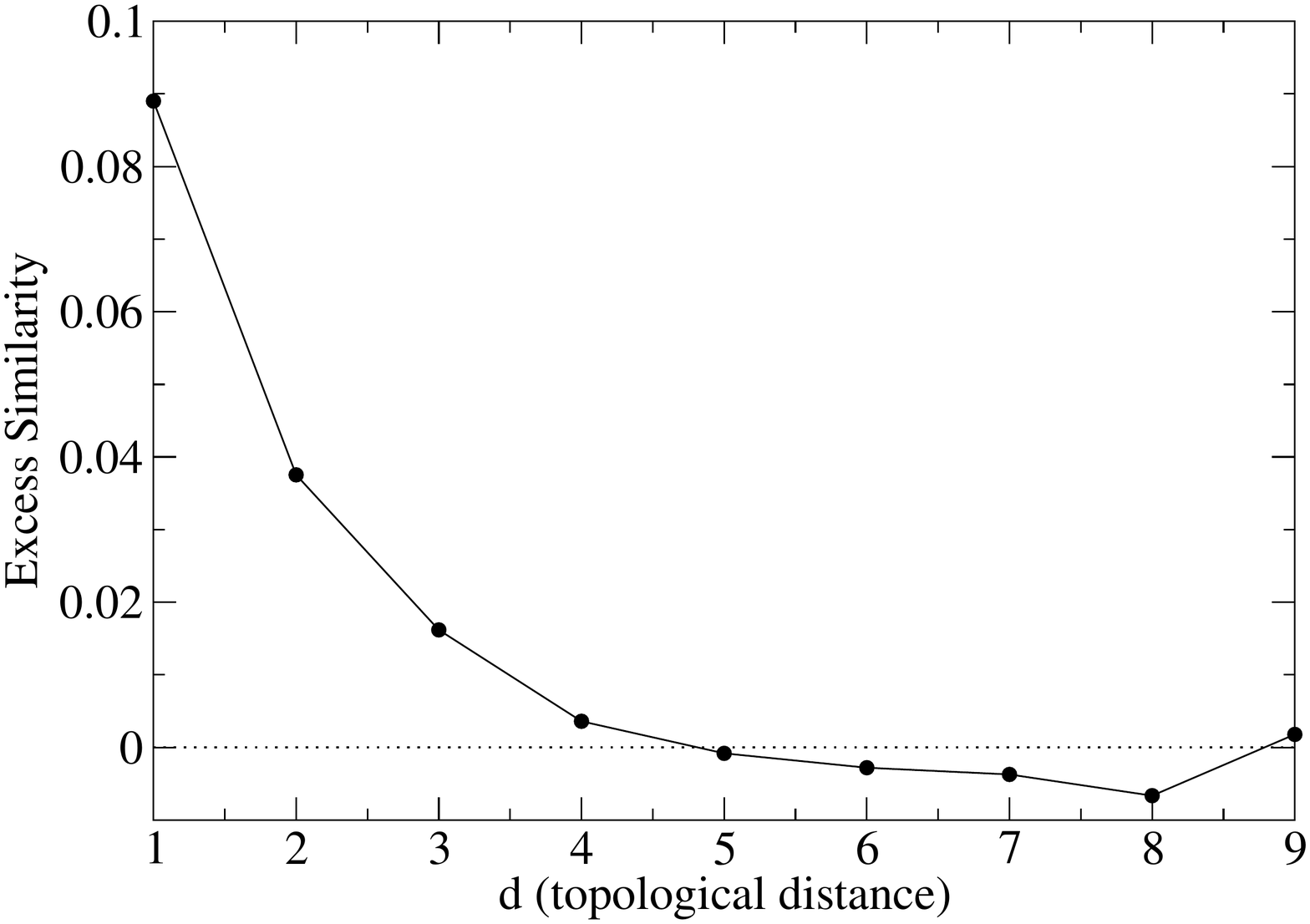}
\caption{{\em Left} The normalized histogram of the lengths for the
  words of the six lexica considered. In the inset, a zoom of the
  region with percentages above the 5\%. {\em Center} The phoneme
  frequency-rank plot for the words of the considered lexica. On the
  x-axis we report the rank of individual phonemes and in the y-axis
  the corresponding frequency in each lexicon. {\em Right} Excess
  Master-Mind similarity of words as a function of the average
  distance $d$ of the corresponding objects on the HBC
  word-association graph.}
\label{Fig:real_languages}
\end{centering}
\end{figure}

First, we measured the distribution of the words length in phonemes
for the lexica of the five languages considered along as for the HBC's
lexicon. The normalized histogram of the lengths it is reported in
Fig.~\ref{Fig:real_languages} (left). For all languages we observe
fairly compatible shapes and in same range we even see identical
trends, it is worth to note that in the most important region (a zoom
of which is portrayed in the inset of Fig.~\ref{Fig:real_languages})
(left) all the languages show different behavior, except for the HBC
and the English language. These similarity was somehow expected since
HBC is actually a sample of the English language.

We now define the frequency of a phoneme in a given language as the
number of words containing that phoneme normalized with the total
number of words in the lexicon of that language. It can also be
roughly considered as the probability of adoption of a given phoneme.
Fig.~\ref{Fig:real_languages} (center) reports the frequency-rank plot
of the phoneme frequencies. After rescaling the frequencies with the
size of the lexicon it is quite evident that the shape of the
distributions is roughly the same, except for the number of different
phonemes in each language corresponding to the maximal rank. The
phonemes frequency has been measured also for the HBC's lexicon. The
figure show that HBC distribution is quite similar to the English
distribution. This fact points out again the reliability of the HBC
database when used to sample the English language properties.

\paragraph{Entropy and Combinatoriality}

A complementary measure of the statistics of use of the different
forms in the lexicon is given by the entropy. The entropy of the
elementary forms distribution is defined as $S=-\sum_{f_i} p(f_i)
\log(p(f_i))$ where $f_i$ is the generic elementary form and $p(f_i)$
is the frequency of occurrence of $f_i$ in the whole emerging lexicon
estimated as $p(f_i)=n(f_i)/F$.

We also introduce a measure of combinatoriality to quantify the
property of a communication system to combine and re-use a small
number of elementary forms to produce a large number of words.
Following the idea in~\cite{galantucci2010}, we introduce a
real-valued quantity ranging in the interval $[0:1]$ that quantifies
how frequently forms recur in a lexicon, according to the following
formula:

\begin{equation}
  C= \frac{\sum_i (m(f_i)-1)}{(M-1) F} \, ,
\label{combinatoriality}
\end{equation}

\noindent where the sum runs over all the $F$ distinct forms present
in the emerged lexicon and $m(f_i)$ is the number of distinct objects
whose name includes the form $f_i$. The term $m(f_i)-1$ takes into
account only the forms that are used to name at least two objects,
i.e., only the forms that are actually re-used. $M$ is the number
words in the lexicon. Table 1 reports the results for the values of
the entropy and combinatoriality for all the lexica considered.

\begin{table}[h]
\label{table_phonemes}
\begin{center}
\begin{tabular}{|c|c|c|c|c|c|}
  \hline 
  word list & $M$ & $F$ & $S$ & $S_{\mathrm{max}}$ & $C$\tabularnewline
  \hline
  \hline 
  \hline 
  British English & 98326 & 77 & 3.574 & 0.8228 & 0.079\tabularnewline
  \hline 
  French & 139721 & 84 & 3.387 & 0.7644 & 0.077\tabularnewline
  \hline 
  German & 327314 & 83 & 3.413 & 0.772 & 0.109\tabularnewline
  \hline 
  Italian & 116878 & 56 & 3.255 & 0.8087 & 0.147\tabularnewline
  \hline
  Spanish & 86016 & 44 & 3.105 & 0.8206 & 0.167\tabularnewline
  \hline 
  \hline 
  HBC & 78954 & 76 & 3.610 & 0.833 & 0.080\tabularnewline
  \hline 
\end{tabular}
\caption{Phoneme statistics of the different languages considered and
  of HBC. $M$ is the number of words in the list, $F$ is the number of
  different phonemes observed in the word list of each dictionary, $S$
  is the entropy calculated on the normalized frequencies
  distributions, $S_{\mathrm{max}}=\log(F)$ is the maximum possible
  value of the entropy, reached when all the forms are equiprobable.
  $C$ is the combinatoriality defined in Eq.~\ref{combinatoriality}.}
\end{center}
\end{table}

\paragraph{Compositionality}

Let us now turn to the compositional aspects of the lexicon, the aim
here being that of establishing whether, in a specific lexicon, words
for semantically related concepts are expressed through
morphologically similar words. To this end we need to focus on a
suitable conceptual graph, i.e., a network where the nodes are the
concepts expressed by words and the connections are semantic relations
between pairs of concepts. Here we shall consider the word-association
graph produced by the HBC experiment as a proxy of the conceptual
space and we shall measure the semantic similarity of two words in
terms of their topological distance on the graph. In addition, we need
to define a measure of morphological similarity between words. To this
end we introduce a Master-Mind-like (MM) measure. Given two words
$w_1$ and $w_2$, each composed of a certain number of forms, the
Master-Mind-like (MM) measure of form similarity is defined as
follows: after the two words have been aligned, either making the
left-end or the right-end coincide, we sum $1$ whenever the two words
share the same form in the same position and $0.5$ for the same form
in different positions. The MM measure will be the maximum between the
left-end and the right-end alignments. The MM measure conveys the idea
that meaningful forms are often included in words as a suffix or a
prefix, and in general in a well defined position.

As a measure of compositionality, we measure the {\em excess
  similarity} of words used to name related objects (at small
topological distance in the graph) when compared to the similarity of
randomly chosen words. In order to do that, we consider the average
difference between the similarity between each pair of words as a
function of their distance on the HBC graph and the same value
computed in the random case, obtained by reshuffling the associations
between words and nodes in the HBC graph.
Figure~\ref{Fig:real_languages} (right) reports the results for HBC.
The excess Master-Mind similarity is stronger for more semantically
related words, i.e. for words a small topological distance in the HBC,
and it monotonically decreases with the topological distance, being
significantly different from zero up to topological distances of the
order of $4$. We interpret this results as a clear signature of
compositionality in HBC and, being HBC a good proxy for English, more
generally for English. Though semantically related words are not
necessarily related in forms, an excess similarity between words used
to name related meanings do exists in natural languages and is
statistically relevant. It would be interesting extending the present
analysis to other languages, for instance by means of suitably
constructed conceptual graphs or proxies for them.

After this short survey of the statistics of phonemes in several
lexica as well as a quantification of the duality of patterning of
these lexica, we now move on a more theoretical ground, by focusing on
a suitable modelling scheme aimed at explaining the emergence of the
above mentioned duality of patterning.

\section{Modelling the emergence of duality of patterning}

We now focus on the mechanisms that could lead to the establishment of
duality of patterning in a lexicon. There have been a number of
previous works devoted to explain the emergence of combinatoriality
and compositionality. A thorough review of the attempts presented in
literature is far from the scope of the present paper. Here we shall
only focus on a few aspects which are relevant for our purposes.

It should be remarked that the two facets of duality of patterning
have often been studied independently from each
other~\cite{nowak_1999anError,
  Oudeyer_2005,debeule_2006,Zuidema_2009}. It should also be remarked
that often studies in this ares have been focused on evolutionary
times scales (e.g., ~\cite{brighton_2002,vogt_2005,kirby_2008}),
disregarding in this way the peer-to-peer negotiation taking place on
cultural time-scales in large populations. In contrast, there is
evidence suggesting that humans are capable of evolving languages with
duality of patterning in the course of only one or two generations
(consider for instance Nicaraguan Sign
Language~\cite{senghas_2004sciencemag} or the emergence of Pidgins and
Creole languages~\cite{mufwene2001ecology}).

Here we aim at explaining in an unitary framework the co-emergence of
combinatoriality and compositionality. In addition, unlike previous
approaches that looked for the emergence of meaning-symbols
compositional mappings out of a small bounded set of predefined
symbols available to the population, our approach adopts an open-ended
set of forms and it does not rely on any predefined relations between
objects/meanings and symbols. For instance we require combinatoriality
to emerge out of a virtually infinite set of forms which are freely
provided to a blank slate of individuals. Such set can only be limited
by means of self-organization through repeated language games, the
only purpose being that of communication. In addition, with our simple
representation of the conceptual space, modeled as a graph, we do not
hypothesize any predefined linguistic category or predefined meaning.
This choice also allows to model the effect of differently shaped
conceptual spaces and of conceptual spaces that may differ from
individual to individual.

\paragraph{The Blending Game}
\label{sec:blending_game}

Here we briefly remind the definition of the Blending
Game~\cite{blending_plos_2012}, where the question of the emergence of
lexicons featuring duality of patterning is addressed in a
self-consistent way (we refer to ~\cite{blending_plos_2012} for the
full definition of the model). We consider an initially {\em blank
  slate} population of $N$ individuals committed to bootstrapping a
lexicon using an open-ended set of forms in a large conceptual space
composed by $M$ objects to be named. We consider those objects as
nodes of a non-directional graph, where links represent conceptual
relations between pairs of objects. All the results we present below
are obtained with a conceptual space modelled as a Erd\H{o}s -R\'enyi
(ER) random graph~\cite{erdos59}. For simplicity, we consider here a
population of agents with an homogeneous structure, where each agent
(individual) has the same probability of interacting with everyone
else.

Starting from scratch and without any central coordination, the
individuals perform pairwise language
games~\cite{steels1995,baronchelli_ng_first,cg_pnas}, where a randomly
selected pair of Speaker (S) and Hearer (H) perform a communication
act aimed at naming objects of their conceptual space. In order to
study the emergence of duality of patterning, we consider the
possibility for any word to be composed either of a single form or of
an ordered linear concatenation of forms\footnote{Here a form is
  intended as the meaningless unit of a signal.}. The Blending Game
features two main ingredients: (i) noise in comprehension and (ii) a
blending repair strategy. Noise is such that H correctly understands
each form uttered by S with a time dependent probability $P_{t}(f) =
\left(1 - \exp\left(-\frac{n_t(f_i)}{\tau}\right)\right)$, where
$n_t(f_i)$ is the number of times H has heard the form $f_i$ up to the
current time $t$, and $\tau$ is a characteristic memory scale. This is
an essential ingredient responsible for keeping the number of
different forms shared by the whole population limited and low,
without any a priori constraints on it. The blending repair strategy
exploits the structure of the world to create new words. Sometimes the
blending is independent of the meaning of these words, feeding the
combinatorial level of the lexicon.

The dynamics of the Blending Game is such that the communicative
success starts from zero and progressively increases, leading
eventually to a fully successful shared communication system with an
``S''-shaped time behaviour. The emerging asymptotic lexicon does not
present homonymy and is such that each object has associated a word
composed by a certain number of elementary forms.

\paragraph{Properties of the emerged lexicon}
\label{sec:properties_lexicon}

\begin{figure}
\begin{centering}
\includegraphics[width=0.32\textwidth]{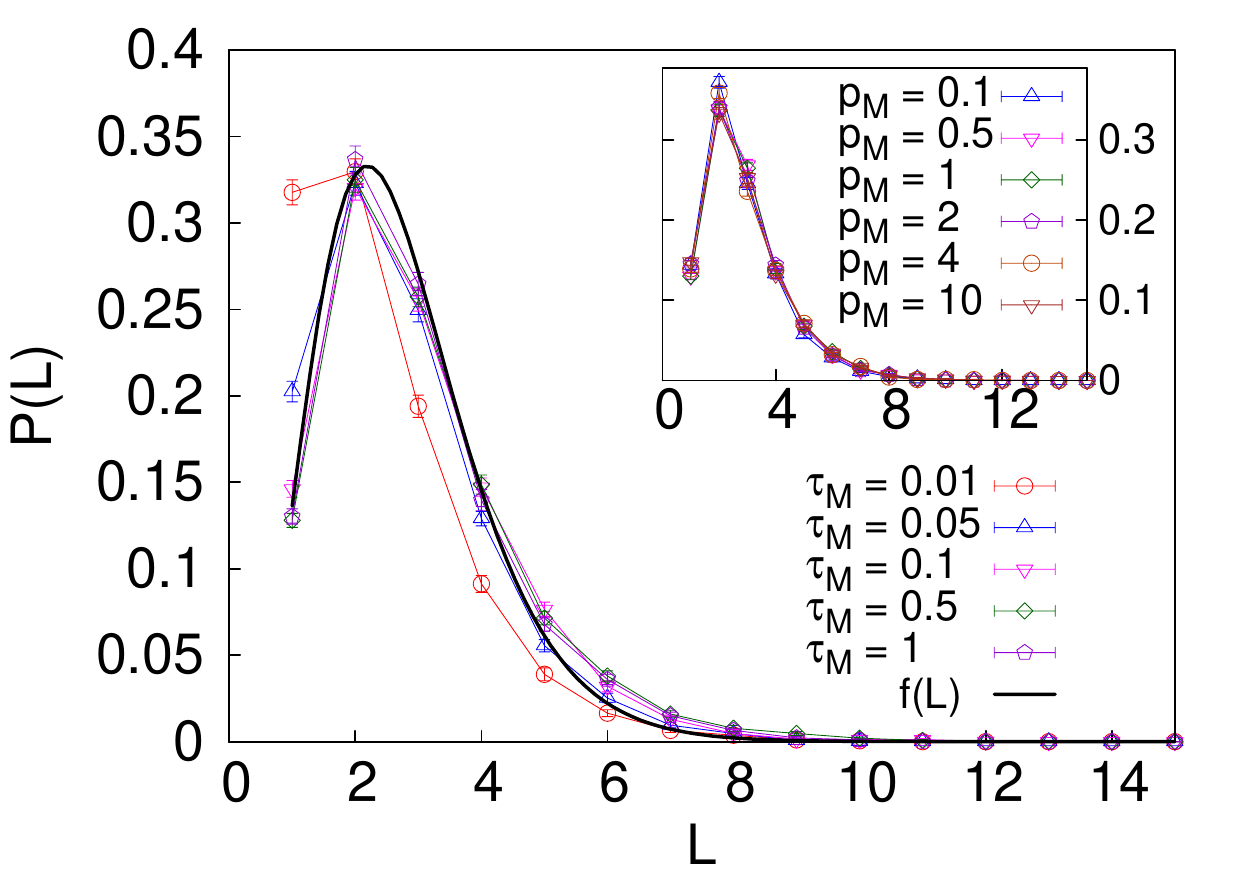}
\includegraphics[width=0.32\textwidth]{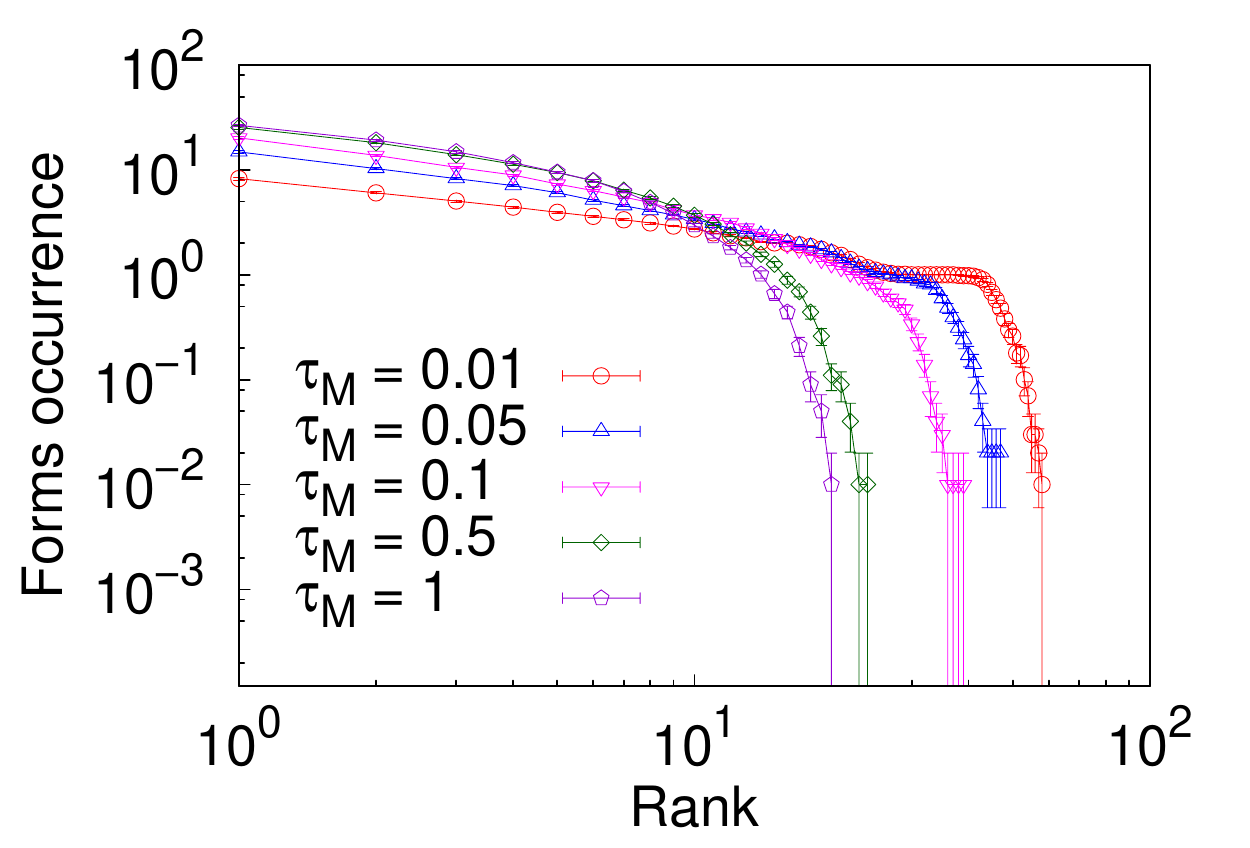}
\includegraphics[width=0.32\textwidth]{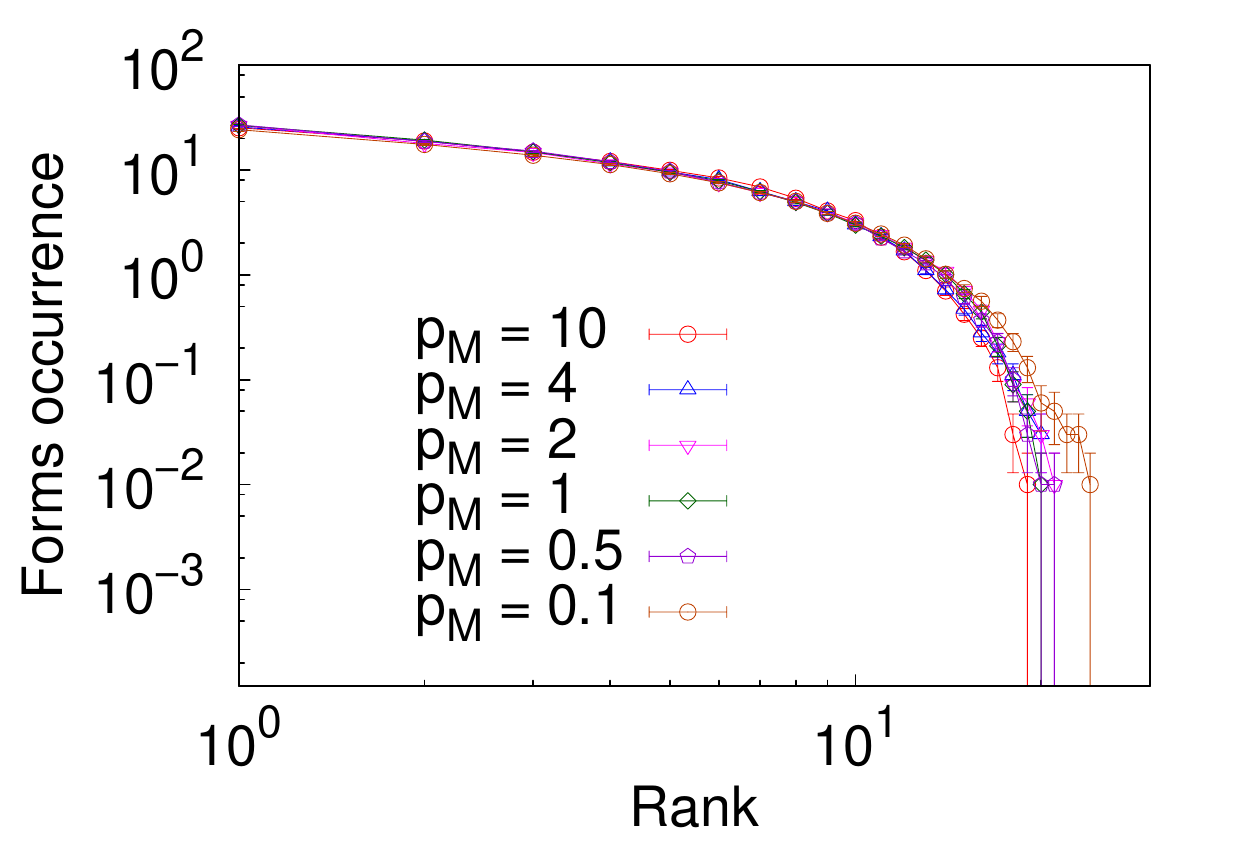}
\vspace{0.1cm}
\caption{{\em Left} Distribution of word length $L$ for different
  values of $\tau$ and $p_{link}$ (the link probability of ER graphs).
  In order to compare graphs corresponding to different sizes, $M$, of
  the conceptual graph, we define the following normalized quantities.
  A normalized link probability for the ER random graph as
  $p_M=\frac{p_{link}}{p^*}$, where $p^*=\frac{\log{M}}{M}$ is the
  threshold above which the whole graph in connected with probability
  one in the infinite size limit ($M \rightarrow +\infty$). Similarly,
  a normalized time scale parameter is defined as
  $\tau_M=\frac{\tau}{M}$. The distribution of word length $P(L)$ in
  the figure is obtained by fixing $p_M=0.5$. $f(L)=a L^b c^L$ is the
  observed empirical function~\cite{Eeg-Olofsson_2004} (solid black
  line, with fitting parameters $a=0.74$, $b=3.7$, $c=0.18$). In the
  top inset we show the same distribution fixing $\tau_M=1$ for
  different values of $p_M$. Note that here the curves overlap,
  indicating that the word length distribution does not depend on the
  objects graph connectivity. The number of agents and the number of
  objects in the environment are fixed respectively to $N=10$ and
  $M=40$. {\em Center} frequency-rank distribution for elementary
  forms is shown for different values of the parameter $\tau_M$
  keeping fixed $p_M=0.5$, $N=10$ and $M=40$. {\em Right} the same
  distribution fixing $\tau_M=1$ and for different $p_M$, showing that
  the distribution of elementary forms does not depend on the objects
  graph connectivity. All the results that follow are averaged over
  different ($100$) realizations of the process on the same conceptual
  graph.}
 \label{fig:model}
\end{centering}
\end{figure}

Due to the blending procedure, the lexicon shared by the population
contains words that are composed by several forms. A typical word in
the lexicon is for instance $f_{23}f_{18}f_0f_0$, composed by the
elementary forms $f_{23}$, $f_{18}$ and $f_0$. We here consider the
length of a word as the number of forms composing it. In
figure~\ref{fig:model} (left), the words length distribution in the
emerging lexicon is reported. We observe that the observed limitation
on the word length is an outcome of the negotiation dynamics, emerging
without any explicit constraints on it. The distribution features a
typical shape that has been observed in human
languages~\cite{Eeg-Olofsson_2004}, well fitted by the function
$f(x)=a x^b c^x$, which corresponds to the Gamma distribution when the
parameters are suitably renamed~\cite{Eeg-Olofsson_2004}.
Figure~\ref{fig:model} (left) shows the word length distribution for
different values of the rescaled memory parameter $\tau_M=\tau/M$ and
of the rescaled graph connectivity $p_{M}$ (see the caption of
figure~\ref{fig:model} for the definition). The average word length in
the emerging lexicon is thus very stable when changing the parameters
of the model, remaining finite and small, and comparable with the
length of words in human languages (compare with
Fig.~\ref{Fig:real_languages}).

\paragraph{Frequency-rank distribution of elementary forms}

As a deeper investigation of the properties of the emerged lexicon, we
consider the frequency-rank distribution of the different forms
composing the words (figure~\ref{fig:model} (center and right)). As in
the case of the word length distribution, the frequency-rank
distribution for forms does not depend on $p_{link}$ (see
figure~\ref{fig:model} right). However, we note in this case a clear
dependence on the memory parameter $\tau_M$. In particular, the higher
$\tau$ (for $M$ fixed), i.e., the lower the fidelity in communication,
the smaller the number of distinct forms on which the agents
eventually find agreement. Since the invention rate of new forms does
not depend on $\tau$, the effect of a high $\tau$ is that of
strengthening the selection process, reducing in this way the number
of forms that get fixed in the population. It is worth noticing that
for large values of $\tau$, the frequency-rank distribution we observe
is remarkably similar to the corresponding distribution observed in
human languages~\cite{tambovtsev_2007} for which a Yule-like
distribution~\cite{yule_1944} has been hypothesized.

\paragraph{Combinatoriality}

\begin{figure}
\begin{centering}
\includegraphics[width=0.49\textwidth]{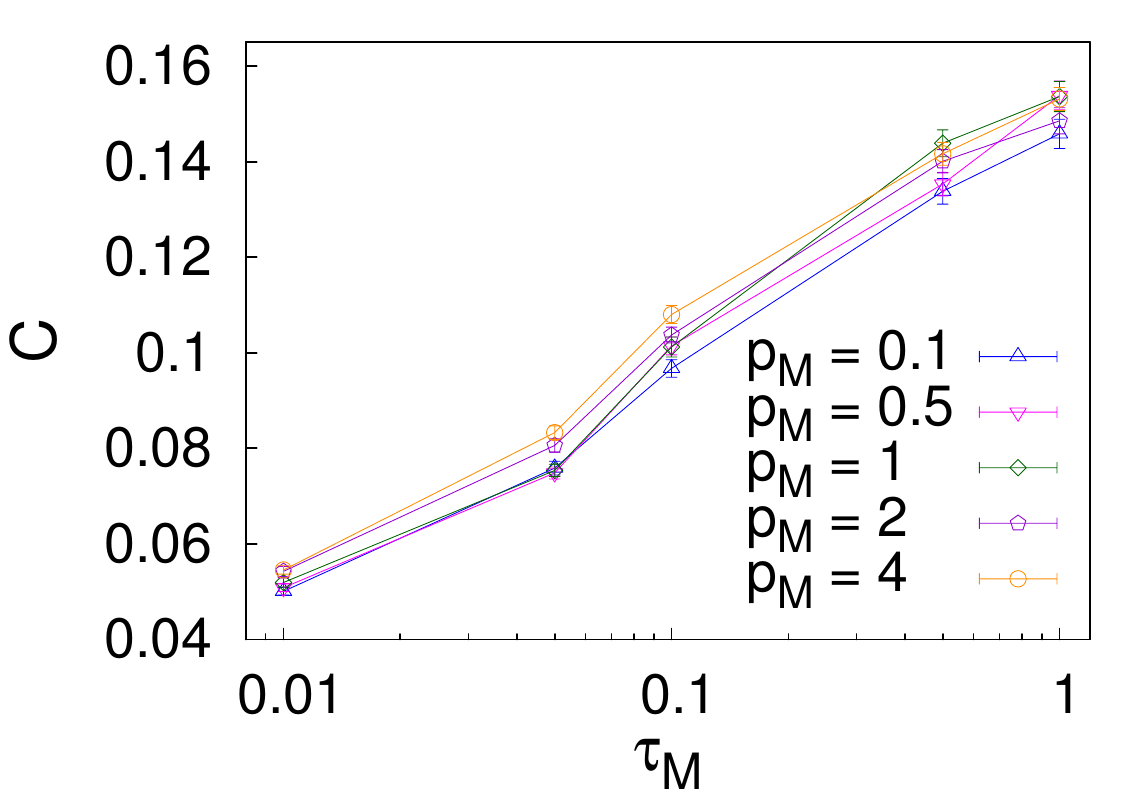}
\includegraphics[width=0.45\textwidth]{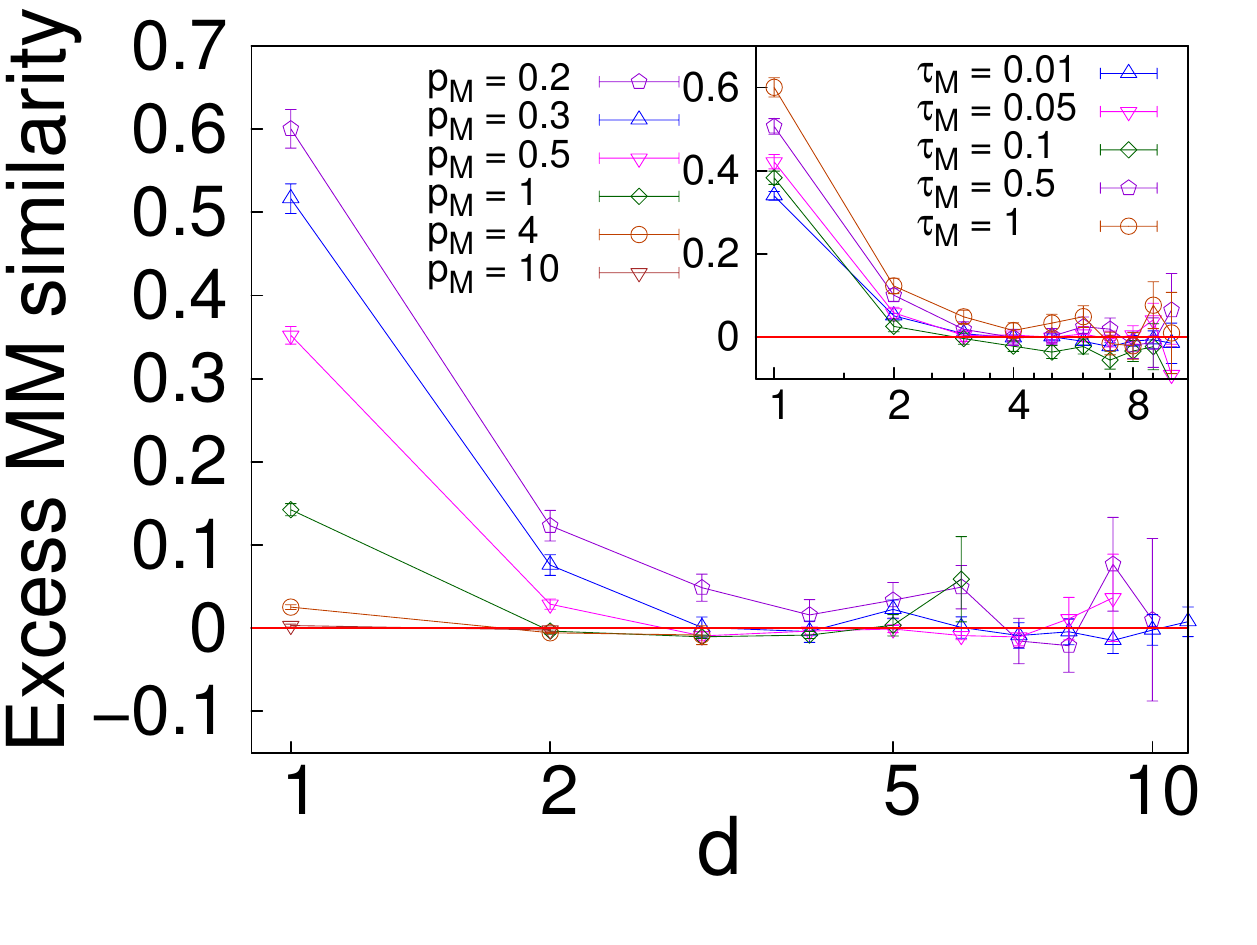}
\vspace{0.05cm}
\caption{{\em Left} Combinatoriality $C$ for different values of $p_M$
  as a function of $\tau_M$. The number of agents and the number of
  objects in the environment are fixed respectively to $N=10$ and
  $M=40$. {\em Right} Excess Master-Mind-like similarity of words as a
  function of the distance $d$ of the corresponding objects on the
  graph. A decrease in the excess similarity as a function of the
  topological distance $d$ is the signature of the emergence of
  compositionality; in particular, compositionality implies higher
  similarity among words which are closer in the semantic space. The
  topological distance on the object graph is our proxy for the
  semantic relatedness. Results are reported for $N=10$ and $M=100$
  and for different values of the objects graph connectivity $p_M$,
  keeping fixed $\tau_M=1$ (main figure) and for different values of
  $\tau_M$ keeping fixed $p_{M}=0.2$ (inset).}
 \label{fig:model2}
\end{centering}
\end{figure}

We now compute the combinatoriality defined in
Eq.~\ref{combinatoriality} to quantify how frequently forms recur in
the emerged lexicon. We report the results in Fig.~\ref{fig:model2}
(left) as a function of $\tau_M$ and for different values of $p_{M}$.
Again, a negligible dependence on $p_{link}$ is found, while, as in
the case of the frequency-rank distribution, a clear dependence on
$\tau_M$ is found, the maximal combinatoriality occurring for high
values of $\tau_M$. This can be understood if one thinks that for a
perfect level of understanding there is no selective pressure acting
on the different forms and many distinct forms are eventually fixed in
the lexicon with a small re-use rate, i.e., little combinatoriality.
Summarizing, when the effort for understanding new forms is
sufficiently high, one finds, at the lexical level, features similar
to the ones observed in real languages, such as the word length
distribution and the number and the frequency of use of the different
forms. In this perspective, combinatoriality emerges as a workaround
to overcome the problem of noisy communication. Interestingly the
values predicted by the Blending Game for the Combinatoriality are not
far from those measured in real languages (see Table 1).

\paragraph{Compositionality}

Let us now turn to the compositional aspects of the lexicon, again
aiming at establishing whether, in the emerged lexicon, words for
semantically related concept are expressed through morphologically
similar words. We refer again to the definition of {\em excess
  similarity} as introduced above. In figure~\ref{fig:model2} (right),
we report the excess similarity for a fixed value of $\tau_M$ and
several values of $p_M$ as a function of the topological distance $d$
on the conceptual graph. Compositionality is evident in the figure:
the more closely related the words, the higher the excess similarity.
It is also remarkable the qualitative similarity with the
corresponding measure performed in real languages (see
Fig.~\ref{Fig:real_languages} (right)). At odds with the above studied
properties of the lexicon, the excess similarity only weakly depends
on $\tau_M$, while strongly depends on $p_M$. This indicates that a
percolation of the organization of the world into the lexicon is
possible when the world has a non-trivial semantic structure, i.e., in
our case when $p_{link}$ is different from zero and from one. In the
former case no relation between objects exists, while in the latter
case all the objects are equally related (all are at distance one in
the graph). Diluted graphs are more prone to induce compositionality.

\section{Discussion and perspectives}

In this paper we have investigated duality of patterning at the
lexicon level. We have quantified in particular the notions of
combinatoriality and compositionality as observed in real languages as
well as in a large-scale dataset produced in the framework of a
web-based word association experiment~\cite{hbc}. We have paralleled
this empirical analysis with a modeling scheme, the Blending Game,
whose aim is that of identifying the main determinants for the
emergence of duality of patterning in language. We analyzed the main
properties of the lexicon emerged from the Blending Game as a function
of the two parameters of the model, the graph connectivity $p_{link}$
and the memory scale $\tau$. We found that properties of the emerging
lexicon related to the combinatoriality, namely the words length
distribution, the frequency of use of the different forms and a
measure for the combinatoriality itself, reflect both qualitatively
and quantitatively the corresponding properties as measured in human
languages, provided that the memory parameter $\tau$ is sufficiently
high, that is that a sufficiently high effort is required in order to
understand and learn brand new forms. Conversely, the compositional
properties of the lexicon are related to the parameter $p_{link}$,
that is a measure of the level of structure of the conceptual graph.
For intermediate and low values of $p_{link}$, semantic relations
between objects are more differentiated with respect to the situation
of a more dense graph, in which every object is related to anyone
else, and compositionality is enhanced. In summary, while the graph
connectivity strongly affects the compositionality of the lexicon,
noise in communication strongly affects the combinatoriality of the
lexicon. 

These results are important because they demonstrate for the first
time that the two sides of duality of patterning can emerge
simultaneously as a consequence of a purely cultural dynamics in a
simulated environment which contains meaningful relations. Many
directions are open for future investigations. First of all to
elucidate the emergence of duality of patterning at the syntax level
beyond that of the lexicon. In addition many different manipulations
of our modelling scheme are possible. One very interesting consists in
relaxing the assumptions that the conceptual space of all the
individuals are identical and modelled as a static graph, imaging
instead that the conceptual space of each individual gets continuously
reshaped by the interactions among the users. In this way one would
realize a truly co-evolution of the conceptual spaces of the
individuals and of their inventories of associations between objects
and words. Finally it is worth mentioning how recent advances in
information and communication technologies allow nowadays the
realization of focused experiments also in the framework of the
emergence of linguistic structures and a general trend is emerging for
the adoption of web-games (see for instance the recently introduced
Experimental Tribe platform: \url{www.xtribe.eu}) as a very
interesting laboratory to run experiments in the social-sciences and
whenever the contribution of human beings is crucially required for
research purposes. This is opening tremendous opportunities to monitor
the emergence of specific linguistic features and their co-evolution
with the structure of out conceptual spaces.

\section{Acknowledgements}
The authors acknowledge support from the KREYON project funded by the Templeton Foundation under contract n.~51663. It is pleasure to warmly thank Bruno Galantucci with whom part of this work has been carried out.


\begin{thebibliography}{10}
\expandafter\ifx\csname url\endcsname\relax
  \def\url#1{\texttt{#1}}\fi
\expandafter\ifx\csname urlprefix\endcsname\relax\def\urlprefix{URL }\fi
\expandafter\ifx\csname href\endcsname\relax
  \def\href#1#2{#2} \def\path#1{#1}\fi

\bibitem{hockett_1960_SA}
C.~F. Hockett, The origin of speech, Scientific American 203 (1960) 88--96.

\bibitem{hbc}
P.~Gravino, V.~D.~P. Servedio, A.~Barrat, V.~Loreto, Complex structures and
  semantics in free word association, Advances in Complex Systems 15 (2012)
  1250054.

\bibitem{galantucci2010}
B.~Galantucci, C.~Kroos, T.~Rhodes, {The effects of rapidity of fading on
  communication systems}, Interaction Studies 11~(1) (2010) 100--111.

\bibitem{nowak_1999anError}
M.~A. Nowak, D.~Krakauer, A.~Dress, An error limit for the evolution of
  language, Proceedings of The Royal Society of London. Series B, Biological
  Sciences 266~(1433) (1999) 2131--2136.

\bibitem{Oudeyer_2005}
P.-Y. Oudeyer, The self-organization of speech sounds, Journal of Theoretical
  Biology 233~(3) (2005) 435--449.

\bibitem{debeule_2006}
J.~D. Beule, B.~K. Bergen, On the emergence of compositionality, in:
  Proceedings of the 6th International Conference on the Evolution of Language,
  2006, pp. 35--42.

\bibitem{Zuidema_2009}
W.~Zuidema, B.~{de Boer}, {The evolution of combinatorial phonology}, Journal
  of Phonetics 37~(2) (2009) 125--144.

\bibitem{brighton_2002}
H.~Brighton, Compositional syntax from cultural transmission, Artificial Life
  8~(1) (2002) 25--54.

\bibitem{vogt_2005}
P.~Vogt, The emergence of compositional structures in perceptually grounded
  language games, Artificial Intelligence 167~(1-2) (2005) 206--242.

\bibitem{kirby_2008}
S.~Kirby, H.~Cornish, K.~Smith, {Cumulative cultural evolution in the
  laboratory: An experimental approach to the origins of structure in human
  language}, Proc. Natl. Acad. Sci. USA 105~(31) (2008) 10681--10686.

\bibitem{senghas_2004sciencemag}
A.~Senghas, S.~Kita, A.~Ozyurek, Children creating core properties of language:
  Evidence from an emerging sign language in nicaragua, Science 305~(5691)
  (2004) 1779--1782.

\bibitem{mufwene2001ecology}
S.~Mufwene, {The ecology of language evolution}, Cambridge University Press,
  2001.

\bibitem{blending_plos_2012}
F.~Tria, B.~Galantucci, V.~Loreto, {Naming a structured world: a cultural route
  to duality of patterning}, PLoS ONE 7 (2012) e37744.

\bibitem{erdos59}
P.~Erd\"os, A.~R\'enyi, On random graphs {I}, Publ. Math. Debrecen 6 (1959)
  290.

\bibitem{steels1995}
L.~Steels, A self-organizing spatial vocabulary., Artificial Life 2~(3) (1995)
  319--332.

\bibitem{baronchelli_ng_first}
A.~Baronchelli, M.~Felici, E.~Caglioti, V.~Loreto, L.~Steels, Sharp transition
  towards shared vocabularies in multi-agent systems, Journal of Statistical
  Mechanics P06014.

\bibitem{cg_pnas}
A.~Puglisi, A.~Baronchelli, V.~Loreto, {Cultural route to the emergence of
  linguistic categories}, Proc. Natl. Acad. Sci. USA 105~(23) (2008) 7936.

\bibitem{Eeg-Olofsson_2004}
B.~Sigurd, M.~Eeg-Olofsson, J.~Van~Weijer, Word length, sentence length and
  frequency - zipf revisited, Studia Linguistica 58~(1) (2004) 37--52.

\bibitem{tambovtsev_2007}
Y.~Tambovtsev, C.~Martindale, Phoneme frequencies follow a yule distribution,
  SKASE Journal of Theoretical Linguistics 4~(2) (2007) 1--11.

\bibitem{yule_1944}
G.~U. Yule, The statistical study of literary vocabulary, Cambridge University
  Press, 1944.

\end{thebibliography}

\end{document}